\documentclass[preprint,12pt]{elsarticle}

 \usepackage{graphics}

\usepackage{color}

\usepackage{amssymb}

\journal{arXiv}

\begin{document}

\begin{frontmatter}



 \title{Balloon-borne gamma-ray telescope with nuclear emulsion : overview and status}





\author[Kobe]{Shigeki Aoki}
\author[Nagoya,Toho_new]{Tsutomu Fukuda}
\author[Nagoya]{Kaname Hamada}
\author[Kobe]{Toshio Hara}
\author[OUS]{Atsushi Iyono}
\author[ISAS,Bern_new]{Jiro Kawada}
\author[Nagoya,Another_new]{Masashi Kazuyama}
\author[AUE]{Koichi Kodama}
\author[Nagoya]{Masahiro Komatsu}
\author[Nagoya,Another_new]{Shinichiro Koshiba}
\author[Nagoya,Another_new]{Hirotaka Kubota}
\author[Nagoya,Tokyo_new]{Seigo Miyamoto}
\author[Nagoya]{Motoaki Miyanishi}
\author[Nagoya]{Kunihiro Morishima}
\author[Nagoya]{Naotaka Naganawa}
\author[Nagoya]{Tatsuhiro Naka}
\author[Nagoya]{Mitsuhiro Nakamura}
\author[Nagoya]{Toshiyuki Nakano}
\author[Nagoya]{Kimio Niwa}
\author[Nagoya,Another_new]{Yoshiaki Nonoyama}
\author[Kobe]{Keita Ozaki}
\author[Kobe]{Hiroki Rokujo}
\author[Nagoya]{Takashi Sako}
\author[Nagoya]{Osamu Sato}
\author[Utsu]{Yoshihiro Sato}
\author[Kobe]{Atsumu Suzuki}
\author[Nagoya]{Kazuya Suzuki}
\author[Nagoya,Kobe_new]{Satoru Takahashi\corref{cor1}}\ead{satoru@radix.h.kobe-u.ac.jp}\cortext[cor1]{Corresponding author}
\author[Utsu]{Ikuo Tezuka}
\author[Nagoya]{Junya Yoshida}
\author[Nagoya,Another_new]{Teppei Yoshioka}

\address[Kobe]{Kobe University, Nada, Kobe 657-8501, Japan}
\address[Nagoya]{Nagoya University, Nagoya 464-8602, Japan}
\address[OUS]{Okayama University of Science, Okayama 700-0005, Japan}
\address[ISAS]{The Institute of Space and Astronautical Science, Sagamihara 229-8501, Japan}
\address[AUE]{Aichi University of Education, Kariya 448-8542, Japan}
\address[Utsu]{Utsunomiya University, Utsunomiya 321-8505, Japan}

\fntext[Toho_new]{Now at Toho University, Funabashi, 448-8510, Japan}
\fntext[Bern_new]{Now at University of Bern, Bern, CH-3012, Switzerland}
\fntext[Tokyo_new]{Now at University of Tokyo, Bunkyou, 113-0032, Japan}
\fntext[Kobe_new]{Now at Kobe University, Nada, Kobe 657-8501, Japan}
\fntext[Another_new]{Now at another institute}

\begin{abstract}
 Detecting the first electron pairs with nuclear emulsion allows a precise measurement of the direction of incident gamma-rays as well as their polarization. With recent innovations in emulsion scanning, emulsion analyzing capability is becoming increasingly powerful. Presently,  we are developing a balloon-borne gamma-ray telescope using nuclear emulsion. An overview and a status of our telescope is given.
\end{abstract}

\begin{keyword}
nuclear emulsion \sep gamma-ray telescope

\end{keyword}

\end{frontmatter}


\section{Introduction}
\label{}
 The observation of high energy cosmic gamma-rays provides us with direct knowledge of high energy phenomena in the universe. The Fermi Gamma-ray Space Telescope with Large Area Telescope (Fermi LAT) was launched in 2008 \cite{FermiLAT}, and large scale observations have been achieved by Fermi LAT since CGRO/EGRET (launched in 1991) \cite{EGRET}. Fruitful results are being obtained in the observation of high energy cosmic gamma-rays. More precise observations will certainly provide more insight into the origin and the mechanism of gamma-ray emission.

 The interaction of high energy gamma-rays with matter is dominated by electron-pair creation processes. The electron pairs contain information of the gamma-ray direction, energy, arrival timing and polarization. By carefully suppressing multiple coulomb scattering and detecting the trajectory of electron pairs, the precise direction of the incident gamma-ray and its polarization can be obtained.

 Nuclear emulsion is a powerful tracking device that can record three-dimensional tracks of charged particles with precise position resolution ($<1\mu$m). Several prominent observations have been performed with nuclear emulsion, i.e. discovery of $\pi$ meson \cite{Powell}, discovery of charmed hadron \cite{Niu} and the first observation of tau-neutrino interactions \cite{NuTau}. In experiments, we use emulsion film (or plate) that has emulsion layers coated on both sides of plastic base. Detecting the first electron pairs with emulsion film allows a precise measurement of the direction of an incident gamma-ray as well as its polarization. We are presently developing a gamma-ray telescope consisting of emulsion films (emulsion gamma-ray telescope).

\section{Emulsion gamma-ray telescope}
\begin{figure}[htbp]
\begin{center}
\includegraphics[scale=0.3,angle=-90]{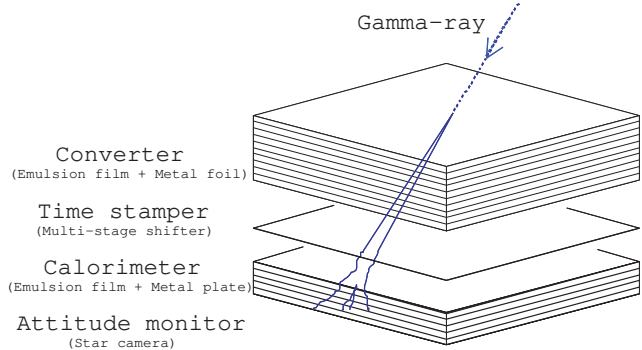}
\caption{The schematic view of emulsion gamma-ray telescope}
\label{EmulsionGammaRayTelescope}
\end{center}
\end{figure}

 Figure \ref{EmulsionGammaRayTelescope} shows a schematic view of the emulsion gamma-ray telescope. The telescope consists of a converter,  time stamper,  calorimeter and an attitude monitor. In the experiment, the telescope must be launched into the stratosphere ($\gtrsim$35km) by balloon. The converter consists of a stack of emulsion films with metal foils. The start of electron pair creation is detected at the converter. The time stamper consists of multi-stage shifter. Multi-stage shifting is new time stamp method for the emulsion, described below. The time stamper gives timing information to converter events, normally lacking in emulsion detection. The calorimeter consists of a stack of emulsion films with metal plates. Gamma-ray energies above several GeV are measured in the calorimeter by measuring the electro-magnetic shower. Gamma-ray energies below several GeV are measured in the converter by measuring multiple coulomb scattering. The attitude monitor consists of a star camera. By combining the attitude monitor information and the event timing, the gamma-ray direction relative to the celestial sphere is determined.

\section{Performance}
 Table \ref{Performance} shows the performance of the emulsion gamma-ray telescope. The telescope has precise angular resolution, no dead time, and it is expected to have polarization sensitivity. Photon statistics from an object is a product of the aperture area and the observation time. The area is limited by emulsion analyzing capability. The time is limited by a chance of the balloon flight and flight duration. For the emulsion analyzing capability, we are developing a high speed automatic emulsion scanning system. With current scanning systems, 2.6 m$^2$$\cdot$flight-time with 100 emulsion films is possible in a year. With future scanning systems, 81 m$^2$$\cdot$flight-time with 100 emulsion films is possible in a year. As the flight duration 150 hours (6.25days), 16 m$^2$$\cdot$day with current scanning systems and 506 m$^2$$\cdot$day with future scanning systems can be obtained, respectively. Our experiment needs to request several long duration flights.

\begin{table}[htbp]
\begin{center}
\caption{The performance of the emulsion gamma-ray telescope}
\begin{tabular}{|l|c|}
\hline Angular resolution @ 100MeV        & 0.57 deg                                \\
\hspace{3.375cm}           @ 1GeV          & 0.08 deg                               \\
\hline Energy range                       & 10 MeV - 100 GeV                    \\
\hline Polarization sensitivity           & Expected                 \\
\hline Aperture area                      & $>$ 1 m$^2$                             \\
\hline Field of view                      & $>$ 2.2 sr   \\
\hline Dead time                          & No dead time                        \\
\hline Area $\times$ Time $^a$ (current$^b$)       & 16 m$^2$$\cdot$day                        \\
\hspace{2.725cm}           (future$^c$)       & 506 m$^2$$\cdot$day                          \\
\hline
\end{tabular}
\\Notes : $^a$ As the flight duration 150 hours, $^b$ Current scanning systems, $^c$ Future scanning systems
\label{Performance}
\end{center}
\end{table}

\subsection{Micro Segment Chamber}
 We started balloon experiments with current emulsion techniques at Sanriku in 2004 \cite{MSC}. The emulsion chamber consisted of a stack of emulsion films and lead plates with shifter mechanism. We call this chamber Micro Segment Chamber (MSC). The shifter had the mechanism to which a part of the emulsion stack shifts relative to other part to allow time resolution of tracks recorded in the emulsion film. The shifter performed to distinguish tracks recorded at each altitude. The MSC performed systematic detection of electro-magnetic component down to 10GeV at level flight by detecting electro-magnetic shower \cite{BS_Nonaka}.

\subsection{Automatic emulsion scanning system}
 We are developing automatic emulsion scanning system \cite{TS}. Currently, five scanning systems are running constantly \citep{SUTS}. Total scanning speed with current scanning systems can be achieved 600cm$^2$/hour. This rate corresponds to 2.6m$^2$$\cdot$100films/year. Development has begun on future scanning system with higher speed. The total scanning speed with two systems under design will achieve 18400cm$^2$/hour. This corresponds to 81m$^2$$\cdot$100films/year. Thus emulsion analyzing capability is becoming increasingly powerful.

\subsection{Angular resolution}
\begin{figure}[htbp]
\begin{center}
\includegraphics[angle=-90,scale=0.3]{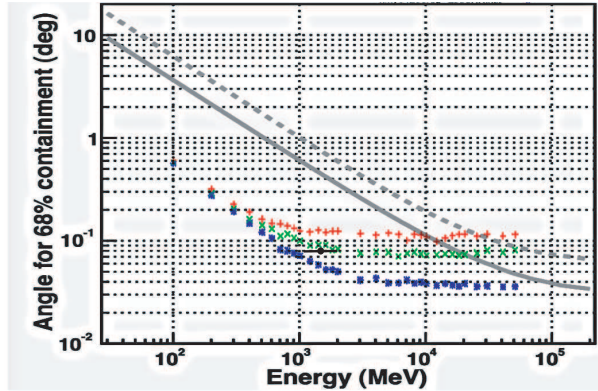}
\caption{The angular resolution of gamma-ray as the function of energy compared with Fermi LAT}
\label{AngularResolution}
\end{center}
\end{figure}

 Figure \ref{AngularResolution} shows the angular resolution of gamma-rays as the function of energy compared with Fermi LAT. Solid line and dashed line show the angular resolution of Fermi LAT. Colored symbols show simulation results of the angular resolution of emulsion gamma-ray telescope for each readout accuracy (+:0.3$\mu$m, $\times$:0.2$\mu$m, $\ast$:0.1$\mu$m). The emulsion gamma-ray telescope is expected to have better angular resolution than Fermi LAT. The dot with error bar shows experimental data using accelerator gamma-ray beam. The angular resolution was confirmed experimentally for this energy region.

\subsection{Energy range}
\begin{figure}[htbp]
\begin{center}
\includegraphics[angle=-90,scale=0.3]{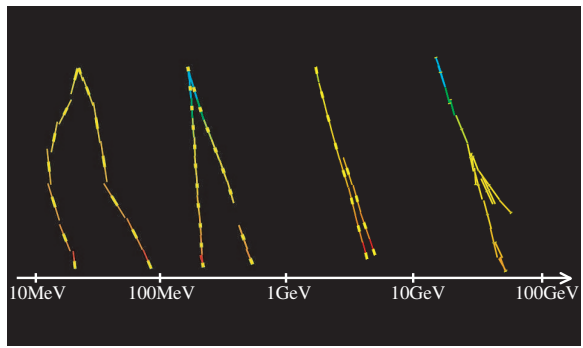}
\caption{Electron pairs at each gamma-ray energy detected with emulsion in test experiments. Chamber structures were, from left to right, a stack of emulsion films, a stack of emulsion films and copper foils (50$\mu$m thickness), a stack of emulsion films and copper foils (50$\mu$m thickness), a stack of emulsion films and lead plates (1mm thickness). Difference of color shows difference of measured film.}
\label{EnergyRange}
\end{center}
\end{figure}

 Systematic detection of electron pairs in emulsion is necessary for the proposed gamma-ray telescope. Test experiments in various energy ranges have confirmed electron pair detection using accelerator gamma-ray beam (SPring-8\footnote{Inverse compton scattering gamma-ray beam, Maximum gamma-ray energy 2.4GeV} and UVSOR\footnote{Inverse compton scattering gamma-ray beam, Maximum gamma-ray energy 47MeV}) and atmospheric gamma-rays at mountain altitude (Mt. Norikura, 2770m) \cite{BSproc}. Systematic detection of electron pairs down to 50MeV was possible in these test experiments. Further study for the detection of electron pairs down to 10MeV is ongoing. Figure \ref{EnergyRange} shows the event topology data of typical electron pairs at each energy detected in these test experiments.

\subsection{Polarization sensitivity}
 A modulation appears for the opening direction of electron pairs created by linearly polarized gamma-rays. By detecting the modulation, polarization can be measured. Emulsion can detect the opening direction of electron pair at its starting point. Therefore, emulsion is expected to have polarization sensitivity by collecting events. We are studying the polarization sensitivity of emulsion using linearly polarized gamma-ray by accelerator.

\subsection{Multi-stage shifter}
\label{MultiStageShifter}
 The attitude of the balloon-borne telescope will change at a rate of approximately a milliradian per second. Thus milliradian angular resolution requires a time resolution less than a second. We developed a multi-stage shifter as the time stamper \cite{MultiStageShifter}. The multi-stage shifter consists of cyclically sliding individual emulsion films relative to each other. By combining track displacement for each stage, many independent states are created. The number of independent states corresponds to the number to resolve the time that the shifter was operating. It is more resoluble for longer time. A multi-stage shifter is similar to an analog clock which shows 12 hours with second accuracy by several hands moving with individual cycles. By increasing number of stages with a shorter cycle, the time resolution is improved. Multi-stage shifter achieves time-stamping by using a simple design, which is compact, light, high- voltage free, low power and without dead-time. We did a test experiment using cosmic rays on the ground to establish the time stamp method by using a multi-stage shifter. A time resolution 1.5 seconds was obtained with two stages.

\section{The flight model}
\subsection{The first flight model}
\begin{figure}[htbp]
\begin{center}
\includegraphics[angle=0,scale=0.175]{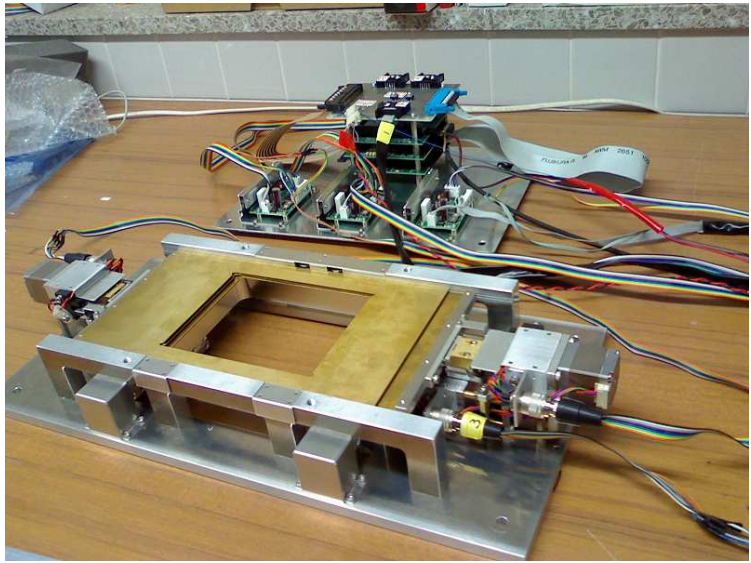}
\caption{The picture of the first flight model}
\label{1st}
\end{center}
\end{figure}

 Figure \ref{1st} shows the flight model of multi-stage shifter\footnote{Co-developed with Mitaka Kohki Co., Ltd.}. The multi-stage shifter has size 44cm$\times$22cm$\times$6cm and a weight of 4.7kg. The central aperture area is 12cm$\times$10cm. The maximum power consumption of the multi-stage shifter is 20W, including the control system shown behind it. The emulsion films is set into the central space, the converter is located above these and the calorimeter is located underneath in the photo, forming a complete emulsion gamma-ray telescope. We tested the first flight model shifter by setting emulsion films under the temperature -25degree and the pressure 1g/cm$^2$ for 12 hours. This multi-stage shifter worked without troubles during the test. Connection accuracy of cosmic ray tracks in emulsion films was also comfirmed without problems, comparable to the test under room temperature and atmospheric pressure. The first flight model is ready for the flight.

\subsection{The second flight model}
\begin{figure}[htbp]
\begin{center}
\includegraphics[angle=0,scale=0.15]{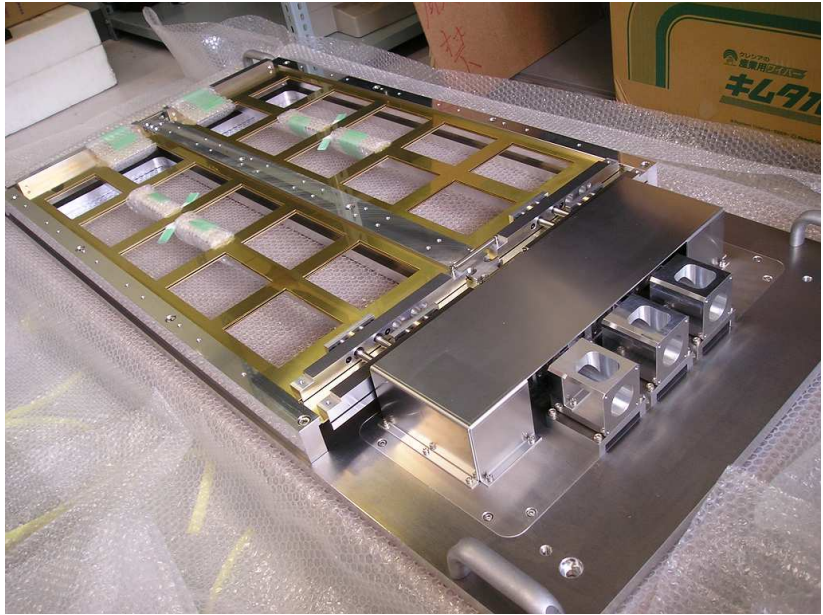}
\caption{The picture of the second flight model}
\label{2nd}
\end{center}
\end{figure}

 Figure \ref{2nd} shows the second flight model expanded the first flight model. The second flight model has size 110cm$\times$64cm$\times$10cm and weight 65kg. Aperture area is 12cm$\times$10cm$\times$20units. Currently, the second flight model is under the operation test and environmental test. The second flight model will be ready for the flight soon.

\subsection{Star camera}
 We are developing a daytime star camera system as the attitude monitor. The design is almost finished. We are considering to use CCD camera consisting of 11.5$\times$13.5$\mu$m$^2$ pixels in a 754$\times$484 array with a transfer rate 30 frames/second and a lens with a diameter 61mm and focal length 85mm (F1.4). This combination gives 0.14$\times$0.16mrad$^2$ as the pixel size and 5.9$\times$4.4deg$^2$ as the field of view projected on the sky. We estimated the performance of considered star camera using Hipparcos star catalog \cite{Hipparcos} and daytime sky background simulated by MODTRAN \cite{StarCamera}. The stellar magnitude limit is 6.8 and an average of 7.2 detectable stars per view were obtained with infrared pass filter ($\geq$750nm). Currently, we are optimizing the design of the hood with baffles to reduce stray light.

\section{Summary and outlook}
 By detecting the first few electron pairs with emulsion, the precise direction of gamma-rays and their polarization can be measured. With the recent innovations in emulsion scanning systems, emulsion analyzing capability is becoming increasingly powerful. By basic study, the perspective was obtained for the observation of cosmic gamma-ray with emulsion. We can begin the observation of cosmic gamma-rays with emulsion gamma-ray telescope. The first flight model is now ready. Using the first flight model, we test under a balloon flight conditions with the flight duration above $\sim$2 hours and measure the background. The second flight model will be ready for the flight soon. With the second flight model, we will test over all by observing known gamma-ray objects with a flight duration above $\sim$6 hours. With a future model, we can start full-scale observations by long duration flights.

\section*{Acknowledgements}
 We would like to thank the people who supported this work in various scenes. Especially, many fruitful comments for the manuscript were given by B. Lundberg. This work was supported by grants from Japan Society for the Promotion of Science, and grants for 21st Century COE Program and Global COE Program from the Ministry of Education, Culture, Sports, Science and Technology of Japan.





\bibliographystyle{elsarticle-num}
\bibliography{<your-bib-database>}



\end{document}